\documentclass[12pt]{article}

\usepackage{graphicx}
\usepackage{amsmath,amssymb}
\usepackage[mathcal]{euscript}

\newcounter{comment}

{\refstepcounter{comment}%
\begin{quote}
\ttfamily\small$\blacksquare$ \textbf{\underline{Comment} $\sharp$\thecomment:}}%
{\end{quote}}

{
\begin{quote}
\ttfamily\small$\blacktriangleright$ \textbf{\underline{Reply} $\sharp$\thecomment:}}%
{\end{quote}}

\newcommand{\cQ}{{\CMcal Q}}

\newcommand{\xBj}{x_{\rm Bj}}
\newcommand{\GeV}{{\rm GeV}}
\font\cmss=cmss12 
\def\1{\hbox{{1}\kern-.25em\hbox{l}}}
\def\bfZ{\relax{\hbox{\cmss Z\kern-.4em Z}}}

\begin{document}

\begin{titlepage}

\centerline{\large \bf A fresh look at exclusive electroproduction of light vector mesons}

\vspace{10mm}

\centerline{\bf M.~Me{\v s}kauskas$^{a}$ and D.~M\"uller$^{a,b}$}

\vspace{8mm}

\centerline{\it $^a$Institut f\"ur Theoretische
Physik II, Ruhr-Universit\"at Bochum} \centerline{\it  D-44780
Bochum, Germany}

\vspace{8mm}
\centerline{\it $^b$Physics Department, Brookhaven National Lab}
\centerline{\it Upton, NY 11973-5000, US}

\vspace{3mm}

\vspace{5mm}

\centerline{\bf Abstract}
\noindent
Relying on the collinear factorization approach, we demonstrate that H1 and ZEUS measurements of exclusive light vector meson and photon electroproduction cross sections can be simultaneously described for photon virtualities of $\cQ\gtrsim 2\, \GeV$. Our findings reveal that quark exchanges are important in this small $x_{\rm Bj}$ region and that in leading order approximation the gluonic skewness ratio is much smaller than one.

\vspace{0.5cm}

\noindent

\vspace*{12mm}


\end{titlepage}

\section{Introduction}
\label{sect-intro}

The  H1 and ZEUS collaborations intensively studied exclusive electroproduction reactions, such as deeply virtual meson production (DVMP) of
$\rho^0$ \cite{Aidetal96a,Breetal98,Adletal99,Breetal99,Adletal02,Aaretal09,Cheetal07}, $\phi$ \cite{Deretal96a,Adletal97,Adletal00a,Cheetal05,Aaretal09}, $\omega$ \cite{Breetal00} and $J/\psi$ \cite{Breetal98,Adletal99a,Cheetal04,Aktetal05} and deeply virtual Compton scattering (DVCS) \cite{Cheetal03,Aktetal05a,Cheetal08,Aaretal09a}, in the small $x_{\rm Bj}$ kinematics. Phenomenologically, the DVMP processes have been widely discussed with color dipole models, see e.g., Refs.~\cite{FraKoeStr95,FraKoeStr97,MarRysTeu96,MarRysTeu99,ForSanSha03,KowMotWat06,MarPesSoy07}, where a rather good description of measurements was reported \cite{Cheetal07,Aaretal09}. The underlying idea of such  models, applicable at small $x_{\rm Bj}$, is that the virtual photon splits into a quark-antiquark pair that interacts via a gluonic $t$-channel exchange with the proton \cite{Muea94,MuePat94,CheMue95,NikZak90,NikZak91a}.  On the other hand a collinear factorization theorem was elaborated that allows to resolve the partonic content in deeply virtual processes by means of perturbation theory \cite{ColFraStr96}. It states that for a longitudinally polarized photon exchange the DVMP amplitudes factorize into mesonic distribution amplitudes (DAs) and generalized parton distributions (GPDs), which are convoluted with a partonic coefficient.
Furthermore, the partonic part, including the changes of GPDs and DAs under scale variation, can be systematically evaluated in perturbation theory and is nowadays known at next-to-leading order (NLO) \cite{BelMue01a,IvaSzyKra04,BelFreMue99}.
Contrarily to the color dipole model, in the collinear factorization approach both quark and gluon $t$-channel exchanges are important for the experimentally accessible  small $x_{\rm Bj}$ region. Also a GPD inspired hand-bag model for the DVMP description has been proposed by Goloskokov and Kroll \cite{GolKro05,GolKro07}. Here, the proton to proton transition is described by a collinear GPD while the parton to meson transition part includes also transverse degrees of freedom.
Although much attention has been given to  H1 and ZEUS  measurements within the aforementioned models, the description of these DVMP data has so far not been explored in the collinear factorization approach.

In the collinear factorization approach the $t$-dependencies of the longitudinally DVMP and DVCS cross sections arise from those of GPDs.  Hence, if these processes would be dominated by gluon exchanges in the small $x_{\rm Bj}$ region,
the $t$-dependencies of the various cross sections should become universal. Experimentally, the exponential $t$-slope of DVMP
cross sections for light vector mesons decreases with growing $\cQ^2$ and  approaches at moderate photon virtuality the DVCS one, however, they are with $B(\cQ^2\sim 4\,\GeV^2)\sim 6/\GeV^2$ larger than the $B^{J/\Psi}\sim 4.5/\GeV^2$ slope  of $J/\Psi$ electroproduction, see, e.g., Fig.~4 in \cite{Lev09}. Taking the universal $t$-dependency criteria literally, it has been argued from the experimental findings that the onset of the perturbative regime appears at rather large photon virtuality of $\cQ^2\sim 15\,\GeV^2$ or so. Theoretically, this is somehow supported by numerical studies in which model dependent NLO corrections turn out to be large \cite{IvaSzyKra04,DieKug07} and, moreover, at this order the residual factorization and renormalization scale dependencies might be still rather strong. Note that these scale setting uncertainties should be maximal at leading order (LO)%
\footnote{At this order, e.g., the ambiguous setting of the factorization scale $\mu$ in the DA and GPD is not compensated by a change of the hard-scattering coefficient, see below (\ref{tffH^pV}).}. On the other hand the DVCS amplitude is in the collinear factorization approach  dominated by quark exchanges rather a gluonic one  and the cross section measurements can be well described at LO and beyond \cite{KumMue09}, where radiative corrections can be considered as moderate \cite{KumMuePas07}. In turn we might argue that DVMP of light vector mesons  in the small $\xBj$ region can be perturbatively described already for a photon virtuality of $\cQ^2 \gtrsim 4\, \GeV^2$, where the $t$-slope deviations of light and heavy meson vector electroproduction cross sections might be attributed to differences in the transverse distribution of sea quarks and gluons.

The most straightforward method to judge on the perturbative description of these processes is a global fit to all of them. For doing so, one needs a flexible GPD parameterization, which is elaborated in terms of a Mellin-Barnes integral transformation that maps conformal GPD moments into the momentum fraction space \cite{MueSch05,KumMuePas07}. Although the NLO corrections are known in this conformal representation \cite{KumMuePas07,LauMuePasSch} the software tools for such DVMP fits are presently under development.
To get a first insight in the phenomenological description of DVMP processes in the small $\xBj$ region by means of the collinear factorization approach,  we
restrict ourselves to the LO approximation and fit flexible GPD models to experimental measurements.

The remainder of the article is as follows.
In Sect.~\ref{sect-model}  we shortly introduce the theoretical formalism and set up our GPD models in terms of conformal moments. In Sect.~\ref{sect-phenomenology}  we confront then this GPD framework with DVMP of  $\rho^0$ and $\phi$ as well as DVCS measurements from the H1 \cite{Aaretal09,Aktetal05a,Aaretal09a} and ZEUS \cite{Cheetal05,Cheetal07,Cheetal03,Cheetal08} collaborations.  We give predictions from  DVCS fits \cite{KumMue10} and the  hand-bag model \cite{GolKro07}, and confront them in return with DVMP and DVCS measurements. We also present the first simultaneous GPD fits to DVCS and DVMP measurements, which illustrate  that in spite of various theoretical and experimental uncertainties  the collinear framework might be applicable in the small $\xBj$ region for $\cQ^2 \gtrsim 4\,\GeV^2 $. Finally, we summarize and conclude.

\section{Formalism and GPD modeling}
\label{sect-model}

The DVCS cross section at small $\xBj$   is dominated by the target helicity conserved  CFF $\CMcal H$:
\begin{equation}
\label{dX-DVCS}
	\frac{d\sigma^{\gamma^*p\to \gamma p}}{dt}\stackrel{\rm Tw-2}{\approx}\pi \alpha^2 \frac{\xBj^2}{\cQ^4}
\left|{\CMcal H}\left(\xBj,t,\cQ^2\right)\right|^2 + \cdots\,,
\end{equation}
where $\alpha\approx 1/137$ is the electromagnetic fine structure constant and the ellipse stays for kinematically suppressed contributions, which
include besides $t/4M^2$ and $\xBj$ proportional terms also non-dominant twist-two CFFs $\widetilde {\CMcal H}$ and $ \widetilde {\CMcal E}$, see, e.g.,
Ref.~\cite{KumMue09}.
To LO accuracy in the running coupling constant $\alpha_s$  the CFF $\CMcal H$ is  decomposed in terms of charge even partonic CFFs, which we denote in the
following as ``CFF''. For four active quarks we write
\begin{equation}
{\CMcal H} = \frac{4}{9} {\CMcal H}^{(u)+} + \frac{1}{9} {\CMcal H}^{(d)+} + \frac{1}{9} {\CMcal H}^{(s)+} + \frac{4}{9} {\CMcal H}^{(c)+}\,,
\end{equation}
where the ``CFFs'' arise from  the convolution of the corresponding GPDs with the LO coefficient,
\begin{eqnarray}
\label{calH^q}
{\CMcal H}^{q(+)}(x_B,t,\mu^2) &\!\!\!\stackrel{\rm LO}{=}\!\!\!& \int_{-1}^1\!dx\,
\left[\frac{1}{\xi-x-i \epsilon} -\frac{1}{\xi+x-i \epsilon}\right] H^{q}(x,\xi,t,\mu^2)\Big|_{\xi=\xBj/(2-\xBj)}\,.
\end{eqnarray}
Here, we express the scaling variable $\xi=\xBj/(2-\xBj)$ by  the Bjorken  variable.

The longitudinally polarized  DVMP cross section reads in the kinematics of interest as follows:
\begin{equation}
\label{dX_L}
	\frac{d\sigma^{\gamma_{\rm L}^*p\to Vp}}{dt}\stackrel{\rm Tw-2}{\approx} 4\pi^2 \alpha\ \frac{\xBj^2}{\cQ^4}
\left|{\mathcal H}^{pV}\left(\xBj,t,\cQ^2\right)\right|^2 + \cdots\,,
\end{equation}
where ${\mathcal H}^{pV}$ is a helicity conserved transition form factor (TFF) and the kinematically suppressed contributions, indicated as  ellipse, include also the target spin-flip TFF ${\mathcal E}^{pV}$ \cite{ManPilWei97a,IvaSzyKra04,DieKug07}.
The dominant TFF ${\mathcal H}^{pV}$ factorizes at leading twist-two and at LO accuracy in $\alpha_s$,
 \begin{eqnarray}
 \label{tffH^pV}
{\mathcal H}^{pV}\left(\xBj,t,\cQ^2\right) \stackrel{\rm LO}{=}
\frac{C_F \alpha_s(\mu_{\rm R})}{N_c} \frac{f_{V}}{\cQ}  3 {\CMcal I}^{V}\left(\mu^2\right) {\CMcal H}^{pV}\left(\xBj,t,\mu^2\right)\,, \quad C_F = 4/3\,, \quad N_C=3\,,
\end{eqnarray}
into the inverse moment of the vector meson DA $\varphi^V(u,\mu^2)$,
\begin{eqnarray}
\label{calI^V}
{\CMcal I}^{V}(\mu^2) = \frac{1}{3}\int_0^1\! du\, \frac{\varphi^V(u,\mu^2)}{u}\,, \qquad \int_0^1\! du\, \varphi^V(u,\mu^2)=1\,,
 \end{eqnarray}
and the  ${\CMcal H}^{pV}$ amplitude that contains the GPDs. Note that the TFF (\ref{tffH^pV}) is proportional to $\alpha_s$ and to $f_V/\cQ$.
Hence, in this approximation the residual renormalization scale $\mu_r$ and factorization scale $\mu$ dependencies are of order $\alpha_s^2$.
For light neutral vector mesons  these ${\CMcal H}^{pV}$ amplitudes  are decomposed as follows
\begin{eqnarray}
\label{cffH^{prho0}}
{\CMcal H}^{p\rho^0} &\!\!\!\stackrel{\rm LO}{=}\!\!\!&
  \frac{1}{\sqrt{2}}
\left(\frac{2}{3} {\CMcal H}^{u(+)}+ \frac{1}{3}{\CMcal H}^{d(+)} +\frac{3}{4}{\CMcal H}^{G}\right),
\\
\label{cffH^{pomega}}
{\CMcal H}^{p\omega} &\!\!\!\stackrel{\rm LO}{=}\!\!\!&
\frac{1}{\sqrt{2}}\left(\frac{2}{3}{\CMcal H}^{u(+)}- \frac{1}{3}{\CMcal H}^{d(+)}+\frac{1}{4}{\CMcal H}^{G}\right),
\\
\label{cffH^{pphi}}
{\CMcal H}^{p\phi} &\!\!\!\stackrel{\rm LO}{=}\!\!\!&  (-1)
\left(\frac{1}{3}{\CMcal H}^{s(+)}+\frac{1}{4}{\CMcal H}^{G}\right),
\end{eqnarray}
where the charge even quark "CFFs" are given in (\ref{calH^q}) and the gluonic one is defined as
\begin{eqnarray}
\label{calH^G}
{\CMcal H}^{G}(x_B,t,\mu^2) &\!\!\!\stackrel{\rm LO}{=}\!\!\!& \int_{-1}^1\!dx\, \frac{1}{2 x}
\left[\frac{1}{\xi-x-i \epsilon} -\frac{1}{\xi+x-i \epsilon}\right] H^G(x,\xi,t,\mu^2)\Big|_{\xi=\xBj/(2-\xBj)}.
\end{eqnarray}
The prefactors in (\ref{cffH^{prho0}}--\ref{cffH^{pphi}}) arise from both
the  electrical charges of quarks and the flavor content of the meson DA.
Furthermore, we take in (\ref{tffH^pV}) for the meson decay constants
$f_{\rho^0} \approx 209\, {\rm MeV}$,  $f_{\omega} \approx 195\, {\rm MeV}$, $f_{\phi} \approx 221\, {\rm MeV}$.

To conveniently treat  evolution,  it is rather popular to expand the DA in terms of conformal partial waves (CPWs).
This provides us for the inverse moment (\ref{calI^V}) the series
\begin{eqnarray}
\label{calI^V-CPWE}
{\CMcal I}^{V}(\mu^2) =
\sum_{k=0 \atop {\rm even}}^\infty E_{k}(\mu^2,\mu_0^2)\, \varphi^{V}_{k}(\mu_0^2)\quad \mbox{with} \quad \varphi_0^{V}  =1\,,
\end{eqnarray}
in terms of CPW amplitudes $\varphi^{V}_{k}$.  Their scale dependency is governed by the evolution operator
\begin{eqnarray}
\label{E_k}
E_k(\mu^2,\mu_0^2)=
\left( \frac{\ln(\mu^2/\Lambda_{\rm QCD}^2)}{\ln(\mu_0^2/\Lambda_{\rm QCD}^2)}\right)^{{\gamma}^{(0)}_{k}/\beta_0},\quad
{\gamma}^{(0)}_{k} =C_F\Big[4 S_1(k+1)-3-\frac{2}{(k+1)(k+2)}\Big], 
\end{eqnarray}
which is defined in terms of the LO anomalous dimensions ${\gamma}^{(0)}_{k}$, where $S_1(k)$ is the first order harmonic sum,
the QCD scale parameter, which  for $n_f=4$ active quarks is set to $\Lambda_{\rm QCD} = 218\,{\rm MeV}$, and the renormalization
coefficient $\beta_0=2n_f/3-11 =-25/3$ of the running coupling.
The inverse moment (\ref{calI^V-CPWE})  is  normalized to one in the asymptotic limit $\mu^2\to\infty$. In the following  we adopt
this asymptotic value also at the input scale. Note, however, that sum-rule estimates  \cite{BalBra96} provide for the $\rho$ meson DA
the second conformal moment
$$a^{\rho}_2(\mu = 1\, {\rm GeV}) = 0.18 \pm 0.1\,,$$ indicating  a moderate deviation from the asymptotic value ${\CMcal I} =1$.

We employ also the CPW expansion  for "CFFs" and GPDs \cite{MueSch05,KumMuePas07,BecMue09}. However, in contrast to
the inverse DA moment (\ref{calI^V}), given as a series (\ref{calI^V-CPWE}), ``CFFs'' (\ref{calH^q},\ref{calH^G}) are now represented in terms of a Mellin--Barnes
integral, which reads in the flavor non-singlet sector as
\begin{eqnarray}
\label{cffH^NS}
{\CMcal H}^{\rm{NS}(+)}
&\!\!\! \stackrel{\rm LO}{=}\!\!\!&
\frac{1}{2 i}\int_{c-i\infty}^{c+i\infty}\!dj\, \xi^{-j-1}
\left[i+\tan\left(\frac{j\pi}{2}\right)\right]  C_j\, E_{j}(\mu^2,\mu_0^2)\,
H^{\rm{NS}(+)}_{j}(\xi,t,\mu^2_0)
\,.
\end{eqnarray}
Here,
\begin{eqnarray}
\left[i+\tan\left(\frac{j\pi}{2}\right)\right]C_j  \quad \mbox{with}\quad
C_j=
\frac{2^{j+1}\ \Gamma\big(\frac{5}{2}+j\big)}{\Gamma\big(\frac{3}{2}\big)\Gamma(3+j)}
\end{eqnarray}
is the hard-scattering amplitude at LO in the charge even sector, the evolution operator  $E_{j}$  is defined in (\ref{E_k}), and
$H^{\rm{NS}(+)}_{j}(\xi,t,\mu^2_0)$ are the conformal GPD moments, analytically continued from the odd ones $j=1,3,\cdots$.
The GPD moments are specified by partial wave amplitudes $H_{j,J}$  that appear in their SO(3) $t$-channel partial wave  expansion
\cite{Pol98},
$$
H_{j}(\eta,t,\mu^2)= \sum_{{J=0 \atop {\rm even}}}^{j+1}  \eta^{j+1-J} H_{j,J}(t,\mu^2)\, \hat{d}^{J} (\eta) \quad\mbox{for}\quad j=1,3,\cdots,
$$
where $\hat{d}^{J} (\eta)$ are (some) Wigner rotation matrices, labeled by $t$-channel angular momentum $J$ and normalized to one in the limit $\eta \to 0$.
In the forward case, $\Delta=0$, the leading SO(3) PW amplitudes, i.e., $J=j+1$, are constrained by the Mellin moments of common parton distribution functions (PDFs),
$$
H_{j}(\eta=0,t=0,\mu^2) = H_{j,j+1}(t=0,\mu^2) = \int_0^1\!dx\, x^j q(x,\mu^2)\,.
$$
In our  GPD model we implement the skewness effect by taking {\em three}  SO(3) partial waves, where the two non-leading ones
are expressed by the leading one, multiplied by the strength $s_{2k}$:
\begin{eqnarray}
\label{H_{jJ}-eff}
H_{j+2k,j+1}(t,\mu^2) = s_{2k}\, H_{j}(t,\mu^2)\,,\quad k=\{0,1,2\}\,,\quad  s_0=1\,.
\end{eqnarray}
We emphasize that this {\em effective} model allows us to control both the normalization of the ``CFF'' and its change under evolution.
Finally, interchanging the $J$ summation with the integration over $j$ provides the formula
\begin{eqnarray}
\label{cffH^NS-model}
{\CMcal H}
&\!\!\! \stackrel{\rm LO}{=}\!\!\!&
\sum_{k=0}^{2}\ \frac{1}{2 i}\int_{c-i\infty}^{c+i\infty}\!dj\, \xi^{-j-1}
\left[i+\tan\left(\frac{j\pi}{2}\right)\right] C_{j+2k}\,  E_{j+2k}(\mu^2,\mu_0^2)\, s_{2k}\,
H_{j}(t,\mu^2_0)
\,,
\end{eqnarray}
which is used for numerical evaluation.
Note that we neglected here the skewness dependency of Wigner`s rotation matrices, appearing in  the CFF (\ref{cffH^NS}), which is a save procedure in the small $\xi$ region \cite{KumMuePas07}.

In the flavor singlet sector the quark combination
$$
{\CMcal H}^\Sigma = \sum_q {\CMcal H}^{q(+)}
$$
and gluon ${\CMcal H}^{\rm G}$ ``CFFs'' will mix with each other.
Adopting  the conventions of Ref.~\cite{KumMuePas07}, we introduce two dimensional  vectors for ``CFFs'' and GPD moments
$$
\mbox{\boldmath{$\CMcal H $}}=\left({ {\CMcal H}^\Sigma  \atop {\CMcal H}^{\rm G}}\right) \quad\mbox{and}\quad
\mbox{\boldmath{$H $}}_{j}=\left({ H_{j}^\Sigma \atop H_{j}^{\rm G}}\right)\,.
$$
In the forward case the moments
\begin{equation}
\left({H_{j}^\Sigma \atop H_{j}^{\rm G}}\right)(t=0,\mu^2)=\int_0^1\!dx\ x^j \left( {\Sigma\atop g}\right) (x,\mu^2)
\end{equation}
are normalized to the   Mellin moments of the common flavor singlet quark ($\Sigma$) and gluon ($G$) PDFs.
Analogously to (\ref{cffH^NS-model}), we model the small $\xi$ behavior of the singlet ``CFF'' as following
\begin{eqnarray}
\label{Res-ImReCFF}
\mbox{\boldmath{$\CMcal H $}}
= \sum_{k=0}^2\frac{1}{2i}\int_{c-i \infty}^{c+ i \infty}\!
dj\,\xi^{-j-1} \left[i +\tan\left(\frac{\pi j}{2}\right)\! \right]
\! \big[
{\mbox{\boldmath ${\mathbb C}$}}\otimes \mbox{\boldmath ${\mathbb E}$}(\mu^2,\mu_0^2)
\big]_{j+2k} \otimes  {\mbox{\boldmath $s$}}_{2k}\otimes
{\mbox{\boldmath $H$}}_{j}(t,\mu^2_0) 
\,,
\end{eqnarray}
where $\otimes$ denotes matrix multiplication. Here,  the LO coefficient matrix reads
\begin{eqnarray}
\label{CPW-LO}
{\mbox{\boldmath ${\mathbb C}$}}_j
\stackrel{\rm LO}{=}
\frac{2^{j+1}\Gamma(j+5/2)}{\Gamma(3/2)\Gamma(j+3)}\left(
  \begin{array}{cc}
 1 & 0 \\
    0 & \frac{2}{j+3} \\
  \end{array}
\right)
\end{eqnarray}
the evolution operator $\mbox{\boldmath ${\mathbb E}$}_j$  is specified as a two dimensional matrix in Sect.~4.2 of Ref.~\cite{KumMuePas07}, and the
two parameters of the diagonal  matrix
\begin{eqnarray}
{\mbox{\boldmath $s$}}_{2 k} =
\left(
 \begin{array}{cc}
 {^\Sigma s}_{2 k} & 0 \\
    0 & {^{\rm G}s}_{2 k} \\
  \end{array}
  \right)
\end{eqnarray}
control the skewness effects of the singlet quark and gluon GPDs.

To have a simple comparison with findings from DVCS fits \cite{KumMue10},
we choose as input scale $\mu^2\approx 4\, {\rm GeV}^2$ and equate renormalization and factorization scales with the photon virtuality.
For charge even contributions,
i.e., nonnegative integer $j$ is odd, the (conformal) moments contain contributions from the valence-, sea-, and anti-quarks. Supposing that the sea quark GPD has the same functional dependence as the anti-quark ones,  we write in analogy to PDF terminology
\begin{eqnarray}
\label{H_j^{q(+)}}
H_j^{q(+)} =  H_j^{q^{\rm val}} + 2  H_j^{\bar q} \approx\,  2  H_j^{\bar q},
\end{eqnarray}
where in the small $\xBj$ region we can safely neglect  valence contributions. Furthermore, we simply assume  that
the functional form is flavor independent, and hence (\ref{H_j^{q(+)}}) can be expressed by the total sea contribution
\begin{eqnarray}
H_j^{q(+)}\approx S^q H_j^{\rm sea},\quad  H_j^{\rm sea}= 2\sum_{\bar{q}=\bar{u},\bar{d},\cdots} H_j^{\bar q},
\quad\mbox{with}\quad
\sum_{q=u,d,\cdots} S^q =1\,,
\end{eqnarray}
where $S^q$ is the flavor asymmetry parameter.  Assuming that charm contributions can  still be neglected at our input scale,
we adopt from global PDF fits a SU(3) flavor asymmetric sea with
\begin{equation}
\label{flavor-scheme}
S^{u}=S^{d}=2S^{s}=\frac{2}{5}
\end{equation}
and we find so at the input scale the following SU(4) flavor nonsinglet  multiplets:
\begin{eqnarray}
\label{H^NS}
H_j^{(3)} &\!\!\!\ =\!\!\!\ &  H_j^{u(+)} -H_j^{d(+)} \approx 0\,,\qquad\qquad
H_j^{(8)} = H^{u(+)} +H_j^{d(+)} -2 H_j^{s(+)} \approx \frac{2}{5} H_j^{\rm sea}\,,
\\
H_j^{(15)} &\!\!\!\ =\!\!\!\ & H_j^{u(+)} +H_j^{d(+)} + H_j^{s(+)} -3 H_j^{c(+)} \approx H_j^{\rm sea}\,,
\nonumber
\end{eqnarray}
which will be evolved autonomously, while as mentioned afore the singlet contribution $H_j^\Sigma\approx H_j^{\rm sea}$ will mix with gluons.
After evolution we finally return to individual quark flavors.

Hence, we have only to model the (conformal) moments $H_j$ for the net sea quark
and gluon contributions at the input scale. In both cases we take for the PDF Mellin moments a simple, however, realistic ansatz  and we decorate it with $t$-dependency
\begin{eqnarray}
\label{H^sea_jj+1}
H_{j}(t,\mu^2) =    N\,
\frac{
B(1-\alpha+j,\beta +1)
}{
B(2-\alpha,\beta+1)}\,
\frac{1+j-\alpha}{1+j-\alpha - \alpha^\prime t }\,
\mbox{\boldmath $\beta$}(t)
\,,\quad \mbox{\boldmath $\beta$}(t=0)=1\,,
\end{eqnarray}
which for $t=0$ corresponds to  a $x^{-\alpha} (1-x)^{\beta}$ PDF  ansatz with momentum fraction average $N$.
Obviously, $\alpha$ and $\beta$ determines the small and large $x$ behavior, respectively. In accordance with phenomenological findings
we fix the $\beta$ parameters to be slightly larger as given by counting rules:
$$
\beta^{\rm sea} = 8\,,\quad \mbox{and}\quad \beta^{\rm G} = 6.
$$
while  $\alpha^{\rm sea} \sim \alpha^{\rm G} \gtrsim 1$ is taken to be the effective intercept of the ``pomeron''.
Note that the momentum sum rule implies the constraint
$$
N^{\rm val} + N^{\rm sea} +   N^{\rm G} =1\,.
$$
In accordance with phenomenological findings from global PDF fits we set the averaged momentum fraction of
$u$ and $d$ valence quarks to $ N^{\rm val}=0.4$  and together with the intercepts for sea quarks and gluons, contained from a simple
PDF fit \cite{KumMue09} to HERA data, we fix the corresponding PDFs at the input scale ${\CMcal Q}_0^2 = 4\, \GeV^2$:
\begin{eqnarray}
\label{PDF-parameters}
N^{\rm sea} = 0.152\,,\quad \alpha^{\rm sea}=1.158 \quad\mbox{and}\quad N^{\rm G} = 0.448\,,\quad\alpha^{\rm G\phantom{a}} &\!\!\! = &\!\!\! 1.247\,.
\end{eqnarray}
The $t$-dependency of the GPD moment (\ref{H^sea_jj+1}) is contained in both the leading ``Regge'' (or ``pomeron'') pole $1/(j+1-\alpha-\alpha^{\prime}t)$  and the residue
\begin{equation}
\label{beta(t)}
\mbox{\boldmath $\beta$}(t) \in \left\{ e^{B\, t}, \left(1-t/m^2\right)^{-2} \right\},
\end{equation}
chosen to be exponential with slope $B$ or as a dipole with cut-off mass $m$. The "pomeron" slope, observed  in  electroproduction processes, is smaller than the soft pomeron one $\alpha^\prime_\mathbb{P}=0.25/\GeV^2$  and we set it here to
\begin{eqnarray}
\label{alpha^prime}
\alpha^{\prime\,{\rm sea}} =\alpha^{\prime\,{\rm G}}  =  0.15/\GeV^2\,.
\end{eqnarray}
The typical value of the slope parameter is for the processes of interest at the input scale  ${\CMcal Q}_0^2 = 4\, \GeV^2$ measured to be
$B =b/2\sim 3\, \GeV^2$  and decreases with growing ${\CMcal Q}^2$. As  noted above, its value for the gluon dominated exclusive $J/\psi$  electroproduction is considerable smaller
$B^{J/\psi}=b^{J/\psi}/2\sim 2.2\, \GeV^2$.  In our fits we  will prefer the dipole ansatz (\ref{beta(t)}) for the residue $\mbox{\boldmath $\beta$}(t)$, where we might take the cut-off masses as in DVCS fits  \cite{KumMue09}
\begin{eqnarray}
\label{M_cut}
M^{\rm sea} = \sqrt{0.5}\,{\rm GeV}\,,\quad  M^{\rm G \ \,} = \sqrt{0.7}\,{\rm GeV}\,.
\end{eqnarray}

\section{Collinear factorization versus measurements}
\label{sect-phenomenology}

\begin{figure}[t]
\begin{center}
\includegraphics[scale=0.45]{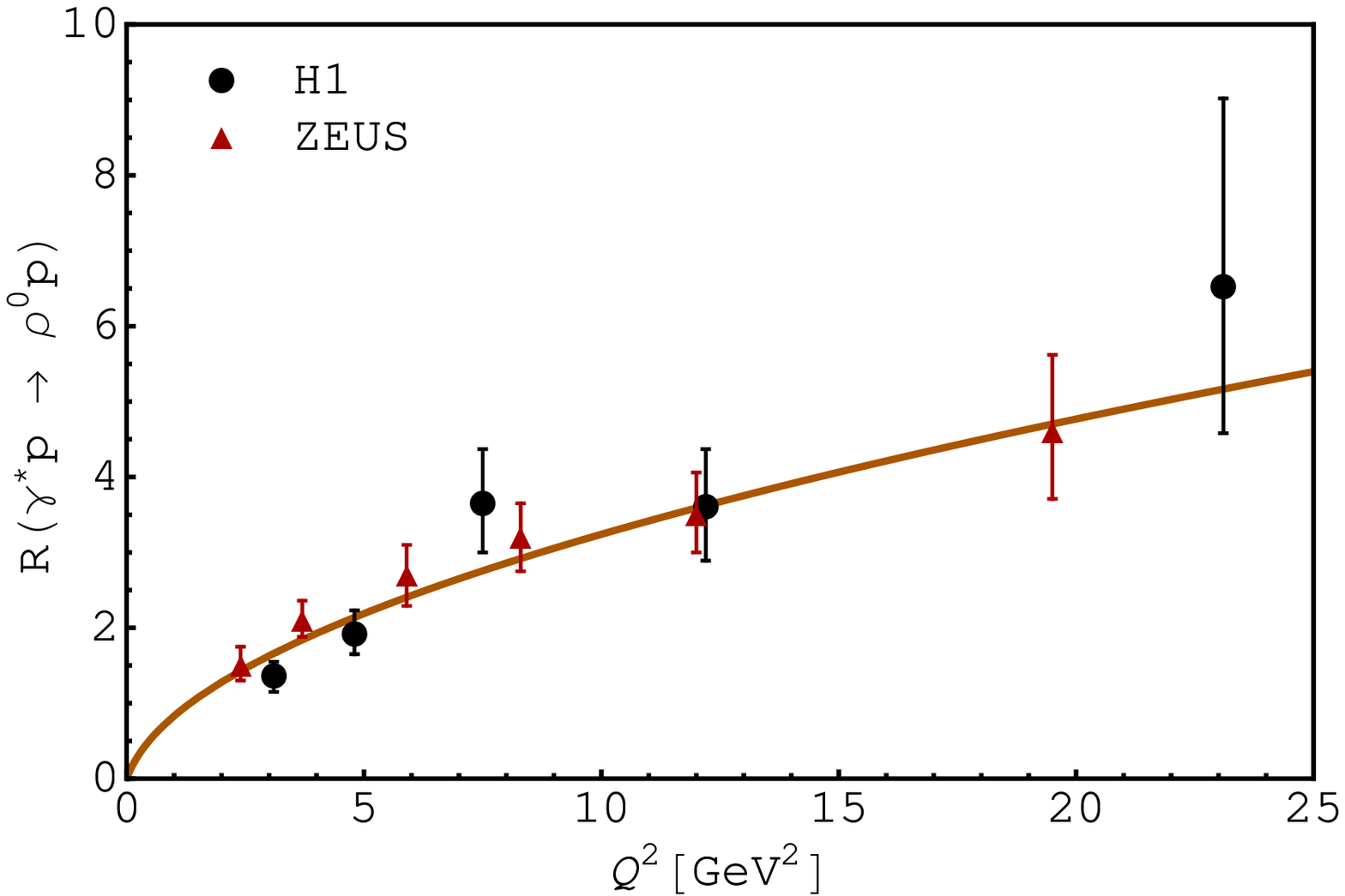}
\hspace{1.cm}
\includegraphics[scale=0.45]{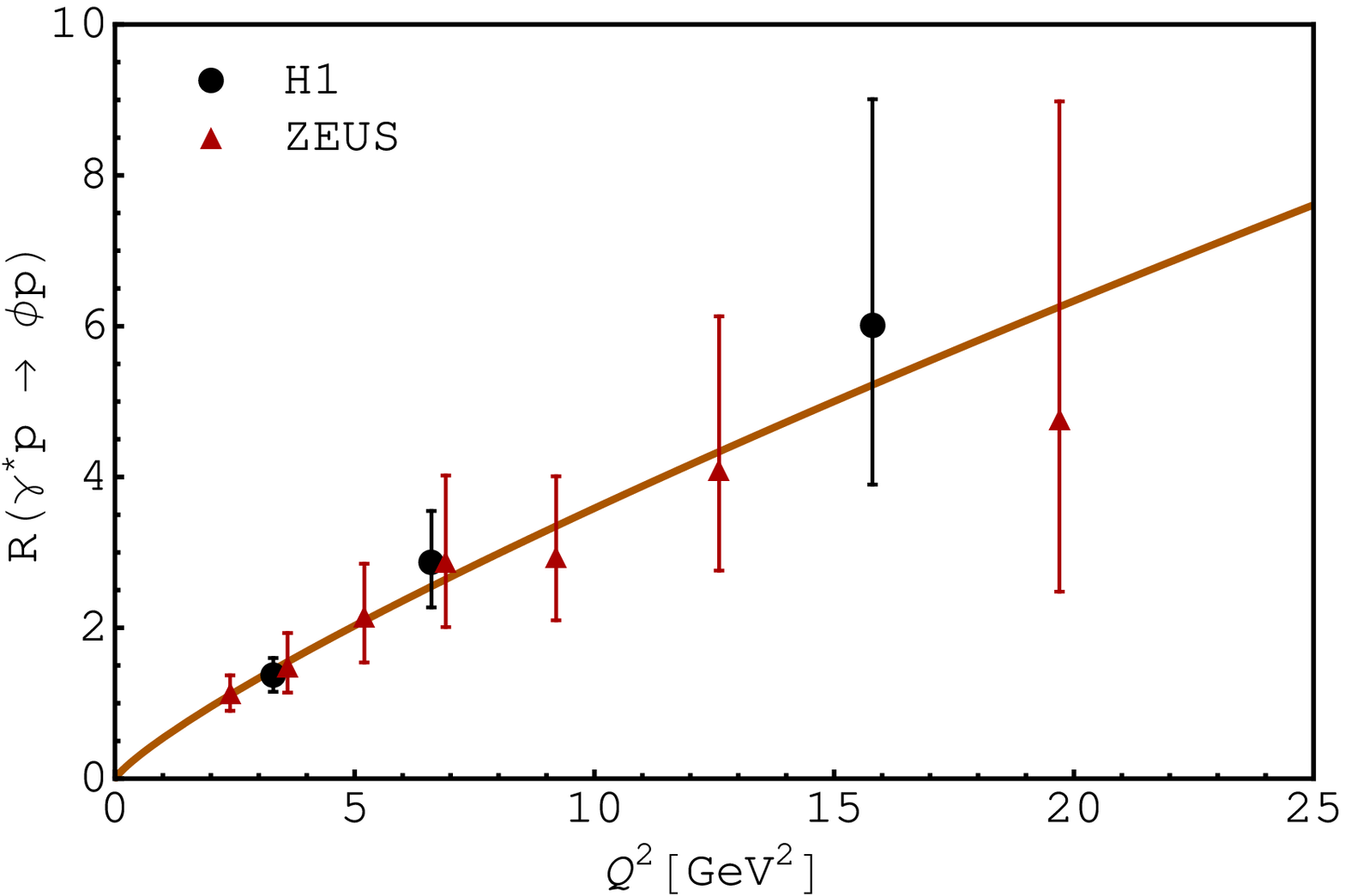}
\vspace{-4pt}
\caption{\small $R$-ratio from the H1 (filled circles) \cite{Aaretal09}  and ZEUS  (filled triangles) \cite{Cheetal07,Cheetal05} collaborations for $\rho^{0}$ (left) and
$\phi$  (right) production, where statistical and systematical errors are added in quadrature.  The solid curves show our fits (\ref{Def-PolPar}).}
\label{fig:R-ratio}
\vspace{-4pt}
\end{center}
\end{figure}

To confront the collinear factorization approach with DVCS and DVMP data,
we evaluate, as described in the previous Section, the differential cross sections (\ref{dX-DVCS}) and (\ref{dX_L})
in terms of our GPD moments (\ref{H_{jJ}-eff},\ref{H^sea_jj+1}). The $t$-integrated cross sections are obtained from the differential ones by
\begin{eqnarray}
\label{X-tintegrated}
\sigma(W,\cQ^2) =\int^{|t_{\rm cut}|}_{|t_{\rm min}|}\!dt\,
\frac{d\sigma(\xBj,t,\cQ^2)}{dt}\Big|_{\xBj= \frac{\cQ^2}{W^2+\cQ^2-M^2}}\,,  \quad M=0.938\, \GeV\,,
\end{eqnarray}
where $|t_{\rm min}|\approx 0$ and $|t_{\rm cut}|\lesssim 1$.
The bulk of DVMP data have been provided by the  H1 and ZEUS collaborations for the total cross sections
$$\sigma= \sigma_{\rm T} +\varepsilon \sigma_{\rm L}, $$
where the photon polarizability
\begin{equation}
\label{Def-PolPar}
\varepsilon\approx
\frac{1-y
}{
1-y+\frac{1}{2}y^2
},\quad 
y= \frac{W^2+\cQ^2 -M^2
}{ s-M^2
}\,, \quad \sqrt{s} = 300\, \GeV\; [{\rm HERAII:}\, 320\, \GeV ]\,.
\end{equation}
To employ in our analysis this larger data set,
we use the experimentally extracted $R=\sigma_{\rm L}/\sigma_{\rm T}$ ratio\footnote{
In the ZEUS analyses  the hypothesis of $s$-channel helicity conservation was employed \cite{Cheetal07,Cheetal05}, while the H1 collaboration used an improved approximation \cite{Aaretal09}. }, in form of a simple fit, shown in Fig.~\ref{fig:R-ratio},
\begin{eqnarray}
\label{R^{exp}}
R^{\rm exp}(\cQ^2)=\frac{\cQ^2/m_V^2}{(1+ a \cQ^2/m_{V}^2)^{p}}\quad \mbox{with} \quad
 \left\{ {a=2.2\,,\; \phantom{1}  p=0.451\,,\; m_V=0.776\,\GeV \atop
 a=25.4\,, \;  p=0.180\,,\; m_V=1.02\,\GeV\phantom{6}} \right\}\;\; \mbox{for} \;\;\left\{ {\rho^0 \atop \phi} \right\},
\end{eqnarray}
where a possible weak $W$ dependency and $t$  dependency is neglected. Although the parameters  $a$ and $p$ are strongly correlated, these fits  indicate that higher twist-effects, parameterized as
$$\frac{1}{R^{\rm exp}} = a^p \left(\frac{m_V^2}{\cQ^2}\right)^{p-1} \left[1+{\CMcal O}(1/\cQ^2) \right], $$ are weaker suppressed than the canonical $1/\cQ^2$ expectation \cite{ColFraStr96}. One might imagine that this modification arises from resumed logarithmical corrections, which are expected from the breakdown of factorization for the transverse polarized cross section \cite{ColFraStr96}. Our predictions, e.g., for the $t$-integrated cross sections (\ref{X-tintegrated})  are then obtained from (\ref{dX_L}) and (\ref{R^{exp}}),
\begin{eqnarray}
\label{X-total}
\sigma(W,\cQ^2)= \left[\varepsilon(W,\cQ^2)+\frac{1}{R^{\rm exp}(\cQ^2)}\right] \int^{|t_{\rm cut}|}_{|t_{\rm min}|}\!dt\,
\frac{d\sigma_{\rm L}(\xBj,t,\cQ^2)}{dt}.
\end{eqnarray}
In utilizing them, we will not take into account the errors from the $R$-ratio fit (\ref{R^{exp}}) and, moreover, as motivated in Sect.~\ref{sect-intro} we will ignore data points at lower photon virtualities, i.e.,  for $\cQ^2 < 4\,\GeV^2 $.

\begin{figure}[t]
\begin{center}
\includegraphics[width=17.2cm]{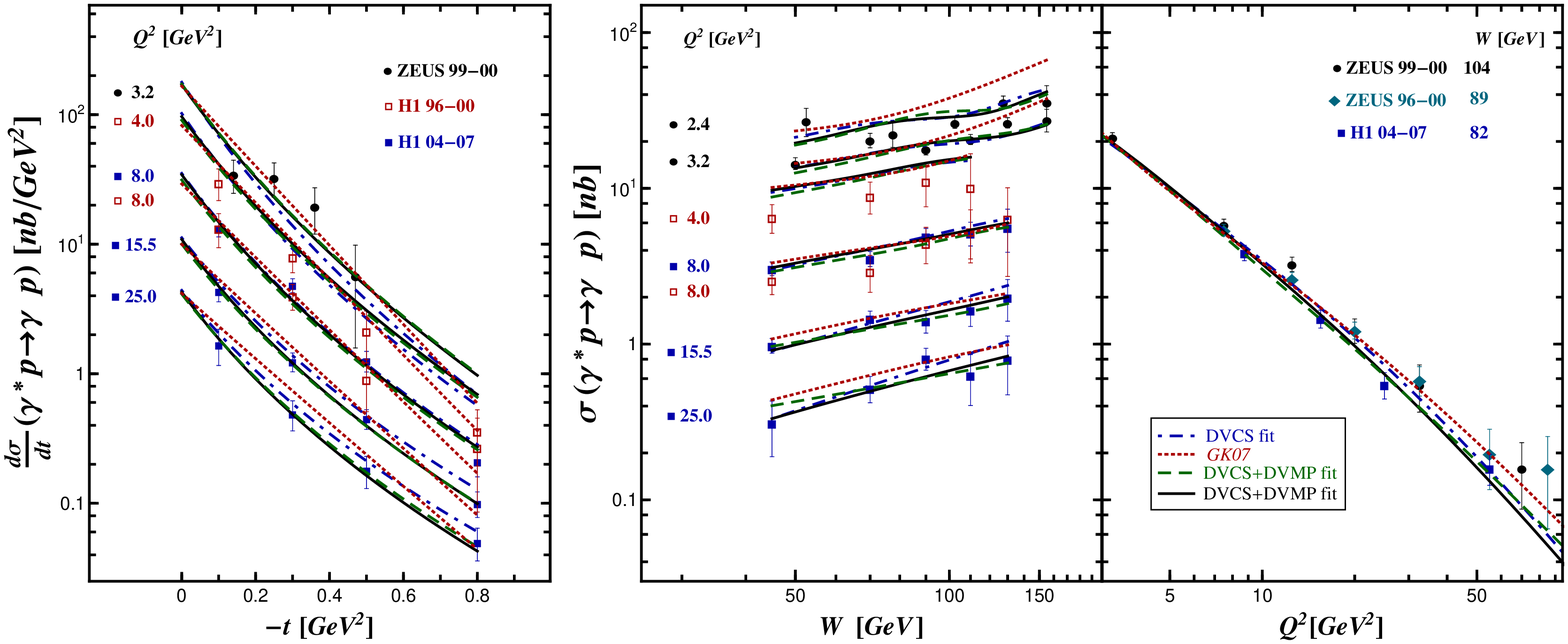}
\end{center}
\caption{\small Differential DVCS cross section  vs.~$-t$ (left)  as well as  $t$-integrated ones vs.~$W$ (middle) and vs.~$\cQ^2$ (right) are taken from \cite{Cheetal03} (filled rhombus), \cite{Aktetal05a} (empty squares),  \cite{Cheetal08} (filled circle), and \cite{Aaretal09a}  (filled squares), where statistical and systematical errors are added in quadrature and normalization uncertainties were ignored.
Measurements are confronted with
a DVCS fit (\ref{fit-DVCS}) (dash-dotted), the {\it GK07} model prediction (dotted) \cite{GolKro07}, and simultaneous DVCS/DVMP fits (\ref{fit-DVCS/DVMP}) (dashed) and (\ref{fit-DVCS/DVMP-H1}) (solid).}
\label{fig:fitsDVCS}
\vspace{-4pt}
\end{figure}
Since we replace here the flavor symmetric sea by the more realistic flavor scheme (\ref{flavor-scheme}) and we include recent DVCS data, we should first update previous DVCS fits \cite{KumMue09,KumMue10}. Thereby, the GPD parameters might be correlated, in particular, this is the case for the two skewness parameters and cut-off mass for both sea quarks and gluons. Note that evolution allows us to access partially the gluon GPD, however, with large uncertainties.   To reach convergency, we use as in previous DVCS fits \cite{KumMue09} the PDF parameters (\ref{PDF-parameters}), "pomeron" slopes (\ref{alpha^prime}), cut-off masses (\ref{M_cut}), and gluonic skewness parameter  $s_4^{\rm G}$ from the {\it KM10b} fit \cite{KumMue10}.
From a three parameter fit to the DVCS cross section
(\ref{dX-DVCS},\ref{X-tintegrated}) measurements of the H1 \cite{Aktetal05a,Aaretal09a} and ZEUS  \cite{Cheetal03,Cheetal08} collaborations  we find
with $\chi^2/{\rm d.o.f.} \approx 130/123$  the new skewness parameters
\begin{eqnarray}
\label{fit-DVCS}
\alpha^{\rm sea}&\!\!\! = & \!\!\!1.158\,, \quad
s_2^{\rm sea} = -0.550\, [-0.460]\,, \quad
s_4^{\rm sea} = +0.130\, [+0.094]\,, \quad M^{\rm sea} = \sqrt{0.5}\,{\rm GeV}\,,
\\
\alpha^{\rm G\phantom{a}} &\!\!\! = &\!\!\! 1.247\,, \quad
s_2^{\rm G\phantom{a}} = -2.397\, [-2.515]\,, \quad
s_4^{\rm G\phantom{a}} = +0.892\, [+0.892]\,,  \quad  M^{\rm G \ \,} = \sqrt{0.7}\,{\rm GeV}\,,
\nonumber
\end{eqnarray}
which slightly differ from the  {\it KM10b} ones, given in square brackets.
The DVCS fit is displayed in the three panels of Fig.~\ref{fig:fitsDVCS} as dash-dotted curves, where the $t$-, $W$-, and $\cQ^2$-dependencies are well described. Note that the  mismatch between the dimensional counting prediction of a $1/\cQ^4$ fall-off for fixed $\xBj$, see cross section (\ref{dX-DVCS}), and the measurements of roughly one power \cite{Cheetal03,Aktetal05a} is resolved by the perturbative prediction of scaling violations. This prediction depends  also  on the chosen parameterization of the non-perturbative distributions at the input scale.
Contrarily to PDFs, where the evolution at small $x$ is essentially determined by the chosen value of the ``pomeron'' intercepts for gluons and the value of the input scale, the GPD evolution is also controlled by the values of skewness parameters $s_2$ and $s_4$ as well as  to some extend by the different $t$-dependencies of flavor singlet quark and gluon GPDs. We also show  predictions from the {\it GK07} model \cite{GolKro07} as dotted curves, where the GPDs were build from Radyushkin`s double distribution ansatz \cite{Rad97} and adjusted to electroproduction data of light vector mesons.
The world DVCS data set for small $\xBj$  is  well described by this  $\chi^2/{n.o.p} \approx 226/126$ prediction, where the model provides an almost perfect LO description and only  fails to describe the $W$-dependency (middle) of the low $\cQ^2 = 2.4 \, \GeV^2$ ZEUS data (filled circles).

\begin{figure}[t]
\begin{center}
\includegraphics[width=17.2cm]{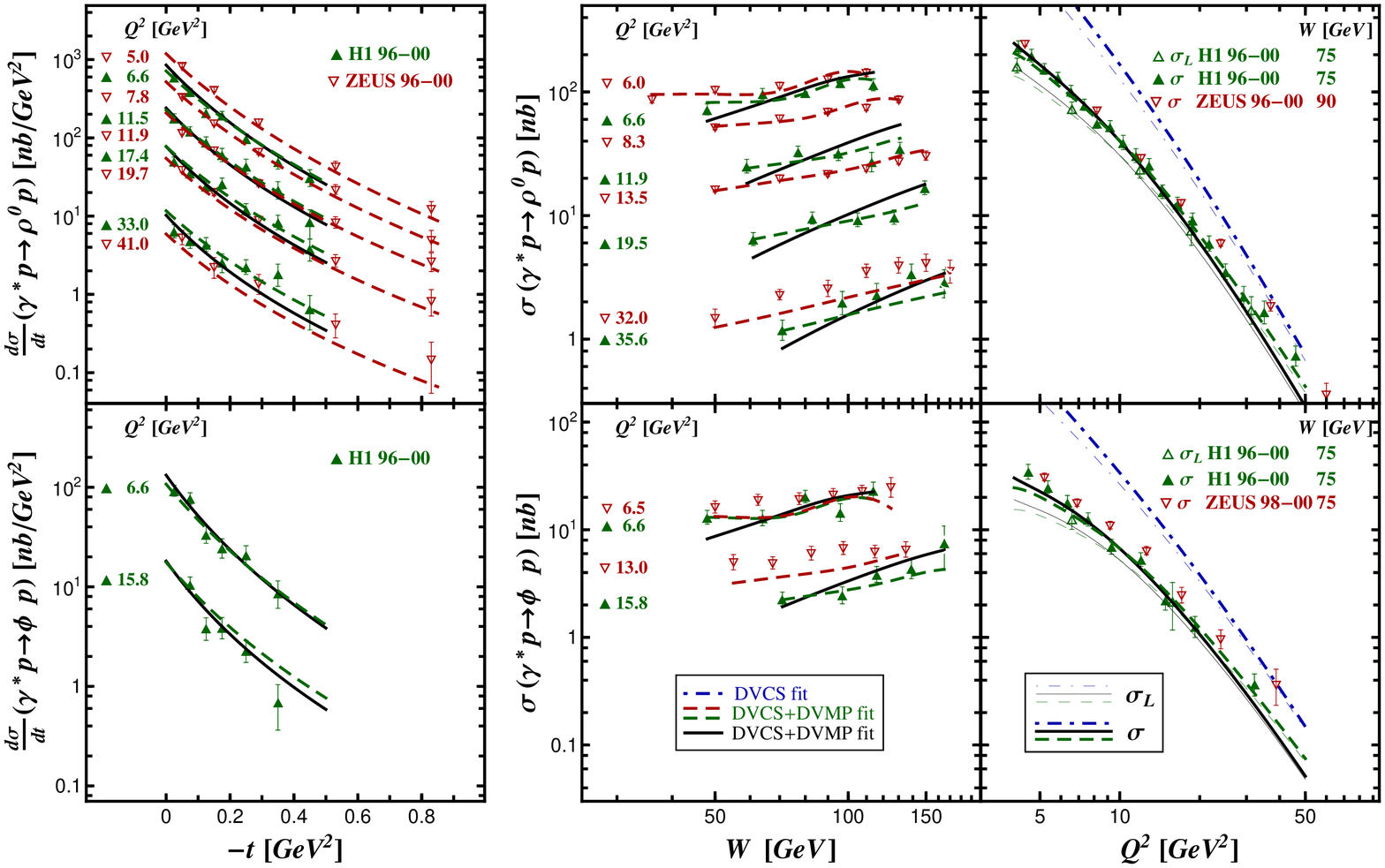}
\end{center}
\caption{\small Differential cross section  versus $-t$ (left)  as well as  $t$-integrated ones versus $W$ (middle) and $\cQ^2$ (right) for DVMP of $\rho^0$  (up) and $\phi$  (down).
Longitudinal (empty up-triangles)  and total cross sections (filled up-triangles)
for $\rho^0$  and $\phi$  meson production form H1 \cite{Aaretal09} and ZEUS (empty down-triangles) \cite{Cheetal07,Cheetal05}.
Statistical and systematical errors are added in quadrature and normalization uncertainties were ignored.
Measurements are confronted with
a DVCS fit (\ref{fit-DVCS}) prediction (dash-dotted) and two  simultaneous DVCS/DVMP fits shown as dashed (\ref{fit-DVCS/DVMP}) and solid (\ref{fit-DVCS/DVMP-H1}) curves.}
\label{fig:fitsDVMP}
\vspace{-4pt}
\end{figure}
A large set of DVMP  data at small $\xBj$ were obtained in the $\rho^0$ channel from the H1 \cite{Aidetal96a,Adletal99,Adletal02,Aaretal09} and ZEUS  \cite{Breetal98,Breetal99,Cheetal07} collaborations ($\phi$ measurements are listed in \cite{Adletal97,Aaretal09,Adletal00a} and \cite{Deretal96a,Cheetal05}). The most recent publications from ZEUS \cite{Cheetal07}  and H1 \cite{Aaretal09} refer to $\rho^0$ data from the 1996/97 and 1998-2000 HERA runs.
The $t$-integrated cross section measurements  for fixed $W=75 (90)\,\GeV$ versus ${\CMcal Q}^2$, where the upper  $t$-cuts are $|t^{\rm ZEUS}_{\rm cut}| = 1\, \GeV^2$  and $|t^{\rm H1}_{\rm cut}| = 0.5\, \GeV^2$, are consistent with each other, see filled up-triangles  \cite{Aaretal09} and empty down-triangles \cite{Cheetal07} on the right upper panel in Fig.~\ref{fig:fitsDVMP}.
The measurement of the longitudinal cross section (empty up-triangles) has been achieved by the H1 collaboration from the knowledge of the spin density matrix elements, shown on the right $\rho^0$ and $\phi$ panels of Fig.~\ref{fig:fitsDVMP}. Thereby, the experimental
errors slightly increase  due to the uncertainties of the $R$-ratio, see Fig.~\ref{fig:R-ratio}.
One also realizes from these panels that our DVCS predictions (dash-dotted curves) overshoot the DVMP cross sections and falls off too steeply with growing ${\CMcal Q}^2$.
Experimentally, the power-like fall-off of the cross sections is determined as $\sim 1/\cQ^4$ \cite{Aaretal09}  while dimensional counting predicts a $1/\cQ^6$ fall-off, see the perturbative prediction (\ref{dX_L},\ref{tffH^pV}).  However, it might be too naively to conclude that these discrepancies already rule out the collinear factorization approach rather they  might be attributed to our relative hard gluon GPD (\ref{fit-DVCS}).

Compared to the  H1 measurements \cite{Aaretal09},  both the $t$- and $W$-dependency  of the ZEUS measurements \cite{Cheetal07} are slightly flatter, e.g.,
\begin{eqnarray}
\left\{{
b^{\rm H1}(\cQ^2=11.5\, \GeV^2)
\atop
b^{\rm ZEUS}(\cQ^2=11\, \GeV^2)}
\right\}
&\!\!\!=\!\!\!&
\left\{
{
6.72 \pm 0.53\;  ^{+0.23}_{-0.25}
\atop
 5.7\phantom{0} \pm 0.5\phantom{0}\;  ^{+0.2\phantom{0}}_{-0.2\phantom{0}}
}\right\}/\GeV^2\,,
\nonumber\\
\left\{{
\delta^{\rm H1}(\cQ^2=6.6\, \GeV^2)
\atop
\delta^{\rm ZEUS}(\cQ^2=6\, \GeV^2)}
\right\} &\!\!\!=\!\!\!&
\left\{
{
0.57 \pm 0.10\phantom{0}\;  ^{+0.05\phantom{0}}_{-0.07\phantom{0}}
\atop
0.4\phantom{0}\pm 0.052\;  ^{+0.048}_{-0.045}
}\right\}\,,
\nonumber
\end{eqnarray}
see also filled up-triangles and empty down-triangles on the upper left and middle panels in Fig.~\ref{fig:fitsDVMP}. A slightly flatter $W$-dependency of the ZEUS data  is also established in the $\phi$ channel \cite{Cheetal05}, see lower middle panel in Fig.~\ref{fig:fitsDVMP}. These differences might be attributed to systematic uncertainties of the background subtractions, in particular of proton dissociative contributions that were experimentally studied by the H1
collaboration \cite{Aaretal09}. Although H1 and ZEUS data are compatible to each other, their separate uses imply some freedom in the partonic interpretation.

The DVMP data from ZEUS \cite{Cheetal07,Cheetal05} are describable to LO accuracy with a very soft gluon GPD, where its leading ``Regge'' intercept  at the input scale $\cQ_0^2=4\,\GeV^2$ is even smaller than one, i.e., $\alpha^{\rm G} < 1$. This ensures that the increase of the effective ``Regge'' intercept with growing $\cQ^2$, which is driven by the $j=0$ pole of the anomalous dimension in the gluon channel, is sufficiently slow. Furthermore, it turns out that  such a very soft gluon GPD is also compatible with the DVMP measurements from H1 and the DVCS data set. This is
illustrated by the dashed curves in Figs.~\ref{fig:fitsDVCS} and  \ref{fig:fitsDVMP}, which arise from a  simultaneous DVCS/DVMP fit with
$\chi^2/{\rm d.o.f.}\approx 618/297$  that pins down the eight parameters
\begin{eqnarray}
\label{fit-DVCS/DVMP}
\alpha^{\rm sea}&\!\!\! = & \!\!\!1.181\,, \quad
s_2^{\rm sea} = 0.565\,, \quad
s_4^{\rm sea} = -0.216\,, \quad
M^{\rm sea} = \sqrt{0.554}\,{\rm GeV}\,,
\\
\alpha^{\rm G\phantom{a}} &\!\!\! = &\!\!\! 0.513\,, \quad
s_2^{\rm G\phantom{a}} = 1.950\,, \quad
s_4^{\rm G\phantom{a}} = -0.469\,,
\quad  M^{\rm G} = \sqrt{0.462}\,{\rm GeV}\,.
\nonumber
\end{eqnarray}
The new sea quark intercept and cut-off mass are entirely consistent with the DVCS fit (\ref{fit-DVCS}). The skewness
parameters have now a reversed sign, providing us an alternative solution to the optimization problem.
However, the very low gluonic intercept might be inconsistent with  PDF findings from deep inelastic scattering (DIS) measurements.
If we fix this intercept, e.g., $\alpha^{\rm G}(\cQ^2=4 \GeV^2)=1.1$, the rather flat $W$-dependency of the ZEUS data implies a mismatch in the normalization. Hence, in such combined H1 and ZEUS DVMP fits we can only get disfavored $\chi^2/{\rm d.o.f.}\sim 5$ values, see also the pomeron fits in \cite{FazFioJenLav11}.

Let us now rely on the DVMP measurements of  the H1 collaboration \cite{Aaretal09}  and DVCS data \cite{Cheetal03,Aktetal05a,Cheetal08,Aaretal09a}.
If we assume a softer gluon PDF with $\alpha^{\rm G}(\cQ^2=4 \GeV^2)=1.1$, as it, e.g., also appears in the hand bag model \cite{GolKro07}, we can reach a  good simultaneous DVMP/DVCS description. Thereby, we adopt the quark PDF from the DVCS fit (\ref{fit-DVCS}) and ask for the remaining six skewness and dipole cut-off parameters. From a $\chi^2/{\rm d.o.f.} \approx 342/224$ fit, shown as solid curves in Figs.~\ref{fig:fitsDVCS} and \ref{fig:fitsDVMP}, we find
\begin{eqnarray}
\label{fit-DVCS/DVMP-H1}
\alpha^{\rm sea}&\!\!\! = & \!\!\!1.158\,, \quad
s_2^{\rm sea} = +0.802\,, \quad
s_4^{\rm sea} = -0.278\,, \quad
M^{\rm sea} = \sqrt{0.548}\,{\rm GeV}\,,
\\
\alpha^{\rm G\phantom{a}} &\!\!\! = &\!\!\! 1.100\,, \quad
s_2^{\rm G\phantom{a}} = -4.288\,, \quad
s_4^{\rm G\phantom{a}} = +1.616\,,
\quad  M^{\rm G} = \sqrt{0.351}\,{\rm GeV}.
\nonumber
\end{eqnarray}
Since the three gluonic parameters $s_2^{\rm G}, s_4^{\rm G}$ and $M^{\rm G}$ are strongly correlated, the $t$-dependency of the gluon GPD can even in this simultaneous DVCS/DVMP fit not be pinned down.

Let us also mention that the  $\omega$ channel might be reasonable described by our two simultaneous DVCS/DVMP fits.
For both of them we find that the longitudinal cross section ratios $\omega/\rho^0\approx 0.1$ at $W=70\, \GeV$  are compatible with the
measured ones \cite{Breetal00}, e.g.,
$$
\frac{\sigma^{\gamma^\ast p \to \omega p}}{\sigma^{\gamma^\ast p \to \rho^0 p}}(W=70\,\GeV,{\cQ^2 =7\, \GeV^2}) = 0.089 \pm 0.014 \pm 0.019 \,.
$$

\begin{figure}[th]
\begin{center}
\includegraphics[width=17.2cm]{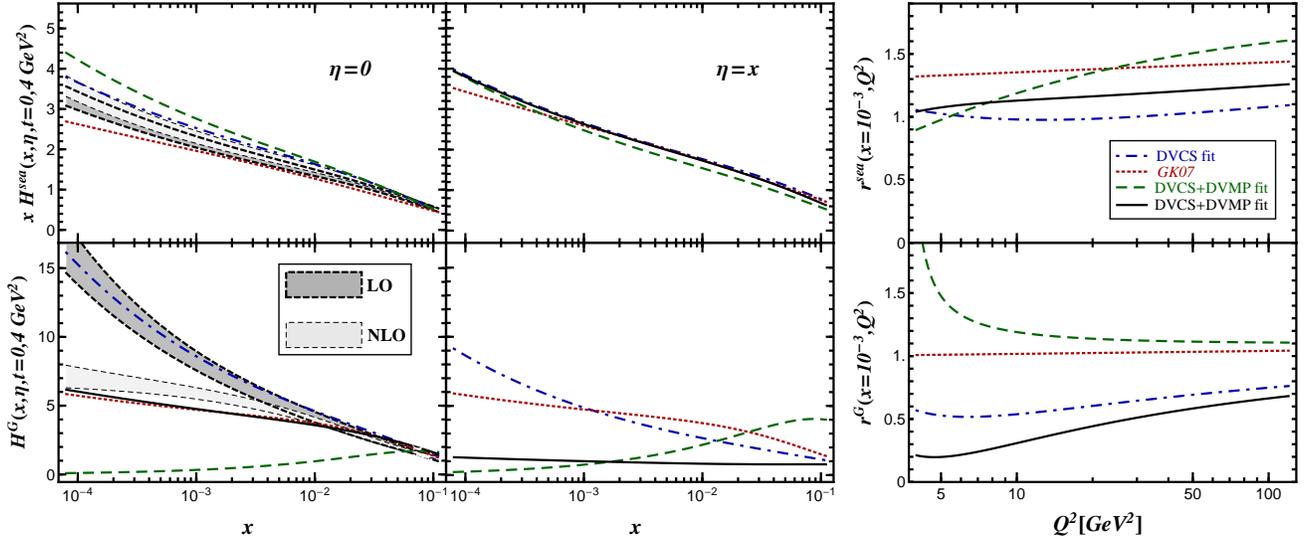}
\end{center}
\vspace{-4pt}
\caption{\small PDFs (left), GPDs on the cross-over line (middle), and the skewness ratios (\ref{r-ratio}) for the flavor singlet sea quark (up) and gluon (down) models, employed in Figs.~\ref{fig:fitsDVCS} and \ref{fig:fitsDVMP}. Phenomenological PDFs at LO (grayed area) and NLO (light grayed area) are taken from Ref.~\cite{Ale02}.}
\label{fig5}
\vspace{-4pt}
\end{figure}
In the left panels of Fig.~\ref{fig5} we compare our models with a standard PDF parameterization of Alekhin \cite{Ale02}. Our total sea quark
PDF from the DVCS fit (dash-dotted curves) is the same as in the simultaneous DVCS/DVMP fit (\ref{fit-DVCS/DVMP-H1}) and it is compatible with Alekhin`s LO parameterization (grayed area). To some extend  this is also the case for the other sea quark models (dashed and dotted curves).  Note also that the NLO radiative corrections in the sea quark sector induce only  mild reparameterization effects, compare grayed and light grayed error bands. The sea quark GPDs on the cross-over line, shown in the upper middle panel, are consistent with each other and so we might conclude that our sea quark models are constrained by DVCS data. Although the skewness parameters in the DVCS (\ref{fit-DVCS}) and simultaneous DVCS/DVMP (\ref{fit-DVCS/DVMP-H1}) fits have different sign and magnitude, the resulting sea quark  GPDs, shown as dash-dotted and solid curves, are hardly to distinguish at the input scale.
Also the gluon PDF in our  DVCS fit (dash-dotted curves) is compatible with standard LO parameterizations.
The gluon PDF of the {\it GK07} model \cite{GolKro07} (dotted curves) and in our simultaneous DVCS/DVMP fit (\ref{fit-DVCS/DVMP-H1}) are rather similar and  underestimate the phenomenological LO findings.
It is worth to mention that radiative corrections will drastically reduce the gluon PDF and, hence, these both aforementioned models are compatible with  NLO PDF parameterizations. The gluon GPD on the cross-over line is for the {\it GK07} model roughly given by the PDF, while in our more flexible models the GPDs on the cross-over line are much smaller than the gluon PDF.  Our simultaneous fit to DVCS and H1/ZEUS DVMP data results in a very soft gluon PDF (dashed curves), which is inconsistent with phenomenological PDF findings.

In the right upper and lower panels of  Fig.~\ref{fig5}  we display the quark and gluon skewness ratios
\begin{equation}
\label{r-ratio}
r^{\rm sea}(x,\cQ^2) = \frac{H^{\rm sea}(x,\eta=x,t=0,\cQ^2)}{2\sum_{q=u,d,s,c} \overline{q}(x,\cQ^2)}\quad\mbox{and}\quad
r^{\rm G}(x,\cQ^2) = \frac{H^{\rm G}(x,\eta=x,t=0,\cQ^2)}{x g(x,\cQ^2)}\,,
\end{equation}
for fixed $x=10^{-3}$ versus $\cQ^2$. For our DVCS (\ref{fit-DVCS}) and simultaneous DVCS/DVMP (\ref{fit-DVCS/DVMP-H1})  fit  we find as previously the value $r^{\rm sea}\approx 1$  \cite{MueKum09}. The sea quarks from these both models, which mostly coincide at the input scale,  evolve only slightly.  The stability of the sea quark ratio under evolution requires that the corresponding gluonic $r$-ratios are smaller than one.
In the collinear factorization framework to LO accuracy  this GPD feature is needed for a successful DVCS fit \cite{MueKum09}.
The sea quark ratio of the {\it GK07} model (dotted curves) is essentially larger, $r^{\rm sea}\approx 1.3$ and rather stable under evolution, too.
Note that the successful DVCS description of the {\it GK07} model presumably originates from the interchange of evolution and skewing procedure \cite{GolKro07}, see numerical examples in \cite{DieKug07a} and comments in Sect.~3.1 of \cite{KumMue09}.

\section{Summary and conclusions}

Based on the collinear factorization approach at LO accuracy and flexible GPD models, we demonstrated that simultaneous GPD fits with $\chi^2/{\rm d.o.f.} \sim 2$ or better describe the kinematical variable dependency of  DVMP (light vector mesons) and DVCS measurements in the small $\xBj$ region already for a photon virtuality of ${\CMcal Q}^2 \gtrsim 4\, \GeV^2$. In our studies we were left with some theoretical and experimental uncertainties. For instance, we did not extensively explore the different partonic degrees of freedom that enter in these processes, in particular, we fixed the flavor content of the quark sea and we relayed on  the asymptotic form of DAs, which do not change under evolution. Moreover, we used naive renormalization and factorization scale setting prescriptions. To utilize the larger data set for the unpolarized DVMP cross sections, we used the experimental $R$-ratio, where the hypothesis of $s$-channel helicity conservation was employed and it was assumed that this ratio only depends on $\cQ^2$. Furthermore, we simplified our analysis by ignoring errors in the $R$-ratio  and normalization uncertainties in the data sets.

In our studies we did not encounter difficulties in the unifying description of $t$-dependency, however, the inclusion of DVMP data from the ZEUS collaboration challenges the expected ``pomeron'' like behavior of the gluon GPD. The successful description of these data  requires a very soft gluon GPD at the input scale of $4\, \GeV^2$, which might be ruled out by standard PDF parameterizations.  Employing only the DVMP measurements from the H1 collaboration together with the DVCS data allow us to describe these exclusive channels, however, with a rather soft gluon GPD that on the cross-over line  is smaller than the gluon PDF. Such a skewness effect at LO accuracy has been already observed in DVCS fits with more realistic gluon PDFs \cite{KumMue09,KumMue10}. Hence, we expect that a reasonable global description of the full  DIS, DVMP and DVCS data set at small $\xBj$ cannot be reached or is disfavored at LO level. On the other hand if we would restrict ourselves to the few released H1 data points for the longitudinally $t$-integrated cross sections, this task might be succeeded. Nevertheless, from our analyses we might conclude that $t$-channel quark exchanges are important in the DVMP processes. While the $t$-dependency of the sea quark GPD is rather well constrained from DVCS data, we certainly realized that the gluon GPD suffers from large uncertainties.

Our partonic interpretation of DVMP differs from those of color dipole models in which by definition the utilized "microscope" is tuned to the gluonic component of the nucleon. Note that the separation of flavor singlet quark and gluon degrees of freedom is implicitly done by adopting a factorization scheme.  In the handbag model approach higher twist contributions, in terms of transverse degrees of freedom,  allows to adopt the popular Radyushkin`s double distribution ansatz for the description of experimental DVMP data. The resulting {\it GK07} GPD model is qualitatively different from our ones. Nevertheless, it also provides a good LO description of DVCS measurements, which originates from the specific modeling of GPD evolution.

Certainly, in our partonic description we are left with some discrepancies between DIS, DVCS, and DVMP findings, which in spite of experimental, theoretical, and model uncertainties cannot be taken as a convincing counter argument against the collinear factorization approach. In our opinion it is worth to study such a global fitting procedure in the NLO approximation of this approach. Thereby, one should also include electroproduction data of $J/\psi$ measurements which strongly constrain the gluon GPD.  The software tool that is needed for this task is under development and, certainly, the fitting procedure should be improved by taking into account the uncertainties of both the $\sigma_L/\sigma_T$ separation and the cross section normalization.

\section{Acknowledgements}
We are grateful to K.~Kumeri{\v c}ki, T.~Lautenschlager, K.~Passek-Kumeri{\v c}ki, and A.~Sch\"afer for many fruitful discussions. D.M.~likes to thank the Nuclear Group at the Brookhaven National Laboratory for the warm hospitality.
This work was supported by a DAAD fellow ship, BMBF grant under the contract no. 06RY9191, and by DFG grant, contract no. 436 KRO 113/11/0-1.


\end{document}